\documentclass[reprint,
superscriptaddress,
amsmath,
amssymb,
 aps,
 pra,
floatfix
]{revtex4-1}
\usepackage{mathrsfs}
\usepackage[cmintegrals]{newtxmath}
\usepackage{graphicx}% Include figure files
\usepackage{dcolumn}% Align table columns on decimal point
\usepackage{hyperref}% add hypertext capabilities
\usepackage[mathlines]{lineno}% Enable numbering of text and display math
\begin{document}
%\preprint{APS/123-QED}
\title{Manipulation of optomechanically induced transparency and absorption by indirectly coupling to an
auxiliary cavity mode}
\author{Guo-qing Qin}
\author{Hong Yang}%
\author{Xuan Mao}%
\author{Jing-wei Wen}%
\author{Min Wang}%
\author{Dong Ruan}%
\email{dongruan@mail.tsinghua.edu.cn}
\affiliation{State Key Laboratory of Low-Dimensional Quantum Physics and Department of Physics, Tsinghua University, Beijing 100084, P.R.China}
\author{Gui-lu Long\textsuperscript{1,}}
\email{gllong@tsinghua.edu.cn}
\affiliation{Beijing Information Science and Technology National Research Center, Beijing 100084, China }
\affiliation{
Beijing Academy of Quantum Information Sciences, Beijing 100193, China}
\date{\today}% 
\begin{abstract}
We theoretically study the  optomechanically induced transparency
(OMIT) and absorption (OMIA) phenomena in a single microcavity
optomechanical system, assisted by an indirectly-coupled auxiliary
cavity mode. We show that the interference effect between the two
optical modes plays an important role and can be used to control
the multiple-pathway induced destructive or constructive
interference effect. The three-pathway interference could induce an
absorption dip within the transparent window in the red sideband
driving regime, while we can switch back and forth between OMIT and
OMIA with the four-pathway interference. The conversion between the
transparency peak and absorption dip can be achieved by tuning the
relative amplitude and phase of the multiple light paths
interference. Our system proposes a new platform to realize
multiple pathways induced transparency and absorption in a single
microcavity and a feasible way for realizing all-optical information processing. 
\end{abstract}
\pacs{Valid PACS appear here}% PACS, the Physics and Astronomy% Classification Scheme.
\maketitle
%\tableofcontents
\section{Introduction}
In the past decades, electromagnetically induced transparency (EIT) has been
studied both theoretically and experimentally  \cite{marangos1998electromagnetically,kash1999ultraslow,fleischhauer2005electromagnetically}. And the potential applications range from ultraslow light
propagation \cite{kash1999ultraslow}, quantum information storage \cite{phillips2001storage,liu2001observation} to the enhancement of nonlinear
processes \citep{harris1990nonlinear,jain1996efficient}. Normally, EIT is a coherent phenomenon due to the destructive quantum interference of two excitation pathways in a
three-level system. On the other hand, the electromagnetically
induced absorption (EIA) \cite{lezama1999electromagnetically,taichenachev1999electromagnetically} is the result of constructive interference between different pathways. Recently, the EIT effects are widely
studied in the optical microresonators \cite{armani2003ultra,jing2018nanoparticle,liu2018gain,liu2017controllable}, i.e. the so called EIT-like effect \cite{totsuka2007slow,yanik2004stopping,xu2006experimental,xiao2009electromagnetically,zhou2011coherent,dong2009modified,peng2014and,wang2016coupled-mode,naweed2005induced,wang2019multiple,xiao2013tunneling,li2016optomechanically,huang2014electromagnetically,yang2015coupled-mode-induced}. Usually, this effect can be generated through a high quality factor (Q) cavity mode directly or indirectly coupling to a low Q one. In this case, a sharp transparent window at their original resonant frequency region appears when the two modes are frequency overlapped.

Based on the strong optical and mechanical interaction, optomechanical systems have potential applications 
in the fundamental research \cite{aspelmeyer2014cavity,pikovski2012probing,liu2018chiral,liu2018observation,chen2017high,cai2017second,xiong2018analysis,abughanem2018cavity,tian2019finite} and provide a promising platform for exploring quantum nonlinear phenomena \cite{jiao2018optomechanical,jiao2016nonlinear}, such as ground state cooling \cite{arcizet2006radiation,marquardt2007quantum,wilson2007theory,chan2011laser,park2009resolved,liu2013dynamic,al2018ground,jing2017high}, entanglement in cavity optomechanical system \cite{wang2013reservoir,palomaki2013entangling,liao2014entangling}. Meanwhile, the cavity optomechanics could enable the exploration of a variety of optical processes. The system has been studied and applied in OMIT \cite{Weis1520,xiong2018fundamentals,kronwald2013optomechanically,safavi2011electromagnetically,lemonde2013nonlinear,kim2015non,shen2016experimental,dong2015brillouin,zhang2018loss,lu2018optomechanically,jing2015optomechanically}, OMIA \cite{safavi2011electromagnetically,qu2013phonon,hocke2012electromechanically}, quantum information processing \cite{stannigel2012optomechanical,dong2012optomechanical,lu2015squeezed} and amplification \cite{nunnenkamp2014quantum,li2017optical,lu2019selective}.

Analog to the EIT (EIA) in the atomic system, the OMIT (OMIA) effect is the result of destructive (constructive) interference between different pathways in optomechanical system. The OMIT effect has been investigated both theoretically and experimentally \cite{Weis1520,xiong2018fundamentals,kronwald2013optomechanically,safavi2011electromagnetically,lemonde2013nonlinear,kim2015non,shen2016experimental,dong2015brillouin,jiang2015chip,lu2017optomechanically}. Compared with OMIT, the OMIA system generally consists of more degrees of freedom or more than two transition pathways for the probe \cite{qu2013phonon,lei2015three,ma2014tunable,jiang2016phase,hou2015optomechanically,bai2016tunable,si2017optomechanically,zhang2017optomechanically,liu2017controllable,ullah2018multiple}. The OMIT could be switched to OMIA through multiple-pathway interference effects, accomplished by coupling to an additional microcavity or adding more mechanical resonators. However, these approaches make the system more complicated and fragile to control.
\begin{figure}[htbp]
\centering
\includegraphics[width=0.9\linewidth]{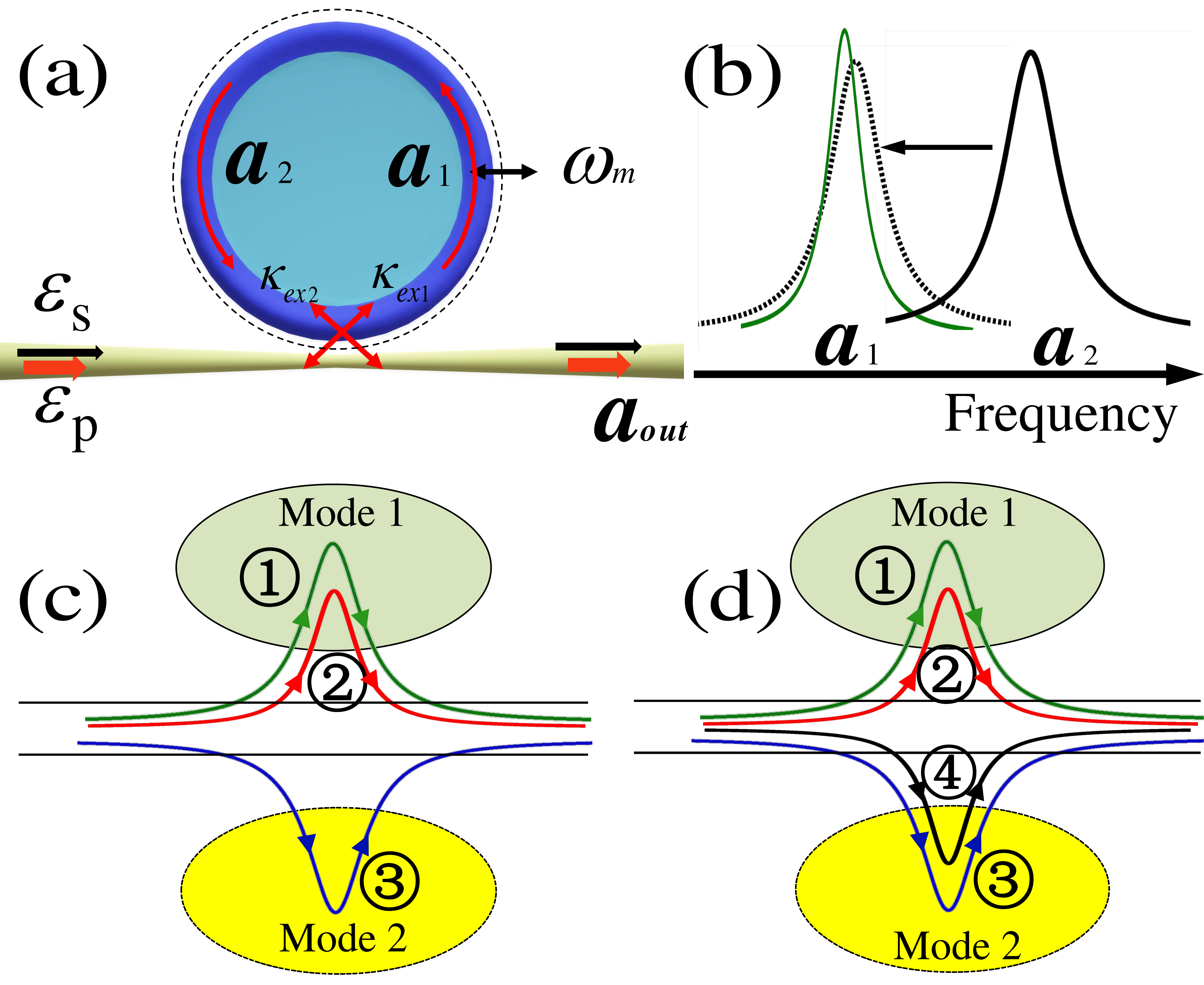}
\caption{(a) Schematic of the optomechanical
system consisting of two optical modes and a mechanical resonator. The two optical modes are excited by the waveguide simultaneously. (b) The auxiliary cavity mode $a_2$ is adjacent to the dominant optical mode $a_1$ in frequency domain. The optical mode $a_2$ can be tuned to overlap with $a_1$, which could lead to electromagnetically-induced transparency-like window\cite{xiao2009electromagnetically}. (c) Illustration of the three-pathway interference effect when the coupling between the auxiliary cavity mode and the mechanical mode can be ignored. (d) Illustration of the four-pathway interference effect when both optical modes couples to the mechanical mode.}
\label{fig1}
\end{figure}

In this paper, we study the optomechanical system and multiple-pathway interference in a single microcavity, in which one cavity optical mode indirectly couples to an auxiliary cavity mode. As shown schematically in Fig.\ref{fig1}, the optomechanical system contains two optical modes and one mechanical mode. The auxiliary cavity mode $a_2$ is adjacent to the dominant optical mode $a_1$ in frequency domain and can be tuned to overlap with $a_1$  \cite{xiao2009electromagnetically}. Different from the directly coupled cavity case \cite{lei2015three,hou2015optomechanically}, the two optical modes here have no direct couple. The interference effect between the two optical modes are mediated by the tapered fiber and could induce the EIT-like phenomenon without the  optomechancial interaction. Combining the EIT-like and OMIT effect, we can easily realize multiple-pathway interference. When the auxiliary cavity mode decouples to the phonons, the three-pathway interference will induce constructive interference effect. We find in the red sideband driving regime, the transmission rate of the system is sensitive to the auxiliary cavity mode. While the auxiliary cavity mode couples to the mechanical mode, the four-pathway interference could induce destructive or constructive interference effect. And the bi-directional transformations between OMIT and OMIA can be achieved. We also study the transmission of the three-pathway interference case in the blue sideband regime. The optical amplification mainly depends on the optomechanical gain and could enhanced by the auxiliary cavity mode. Therefore, the output spectrum can be precisely controlled by tuning the auxiliary cavity mode and the optomechanical coupling strength.
The paper is organized as follows. In Sec~\ref{section2}, we introduce
the theoretical model of the
optomechanical
system shown in Fig.\ref{fig1}. In Sec~\ref{section:3}, we study the physical origins of the three light propagation paths when the auxiliary cavity mode decouples to the mechanical mode. And we discuss the optical response of probe field both in red and blue sideband driving regime. In Sec~\ref{section:4}, the bi-directional conversion between OMIT and OMIA can be achieved through the four-pathway interference effect. In Sec~\ref{section:5}, the conclusion
is given.
\section{Model}
\label{section2}
As schematically shown in Fig.\ref{fig1}, the system we considered here is a whispering-gallery-mode optomechanical microresonator, where two optical modes and one mechanical mode have been excited. The dominant optical mode $a_1$ has the resonant frequency $\omega_1$ and the internal loss $\kappa_{10}$.  The auxiliary cavity mode $a_2$ has the resonant frequency $\omega_2$ and the internal loss $\kappa_{20}$. 
The mechanical resonator has the frequency $\omega_m$ with the effective mass $m$.  The optical modes $a_1$ and $a_2$ are two independent WGMs. However, the cavity mode $a_1$ can couple to the mode $a_2$ indirectly through the waveguide \cite{xiao2009electromagnetically}. The interference between the two optical pathways results in EIT-like line shape. Both optical modes could couple to the mechanical resonator with different optomechanical coupling rate $g_1$ and $g_2$. As shown in Fig.\ref{fig1} (c) and (d), the optomechanical coupling strength between the auxiliary cavity mode and the mechancial mode is weak. When the optomechanical coupling $g_2$ can be ignored, the system shows three pathways interference. Without loss of generality, the system shows four pathways interference when both optical modes couple to the mechanical mode. The Hamiltonian can be described as  
\begin{align}
H &= \hbar \omega_1 a^\dagger_1 a_1+\hbar\omega_2 a^\dagger_2 a_2+\frac{1}{2}m\omega^2_m x^2+\frac{P^2}{2m}\nonumber\\
& +\int_{-\infty}^{+\infty} \hbar \omega c^\dagger(\omega)c(\omega)\, d\omega+\hbar g_1 x a^\dagger_1 a_1 +\hbar g_2 x a^\dagger_2 a_2\nonumber\\
& +i\hbar \sum_{j=1,2} \kappa_{ex,j}(\omega)[c^\dagger(\omega)a_j-a^\dagger_j c(\omega)]\,d\omega\label{1}
\end{align} 
where the operators $x$ and $P$ represent the position and momentum
of the mechanical mode, respectively. The annihilation operator
$c(\omega)$ indicates the waveguide mode, which satisfies $[c(\omega),c^\dagger(\omega^{'}) ]=\delta(\omega-\omega^{'})$. 
$\kappa_{ex,j}(\omega)$ (j=1,2) describes the coupling constant
between the cavity mode $a_j$ and waveguide modes. The
first four terms represent the free Hamiltonians of the optical and mechanical modes. The fifth term is to
describe the Hamiltonian of the waveguide modes. And the last term indicates the the coupling between the optical modes and the waveguide modes. In our scheme, the system is driven by a strong control laser field
with the amplitudes $\varepsilon_p$ and frequency
$\omega_p$, respectively. Meanwhile, a weak probe laser field, with the amplitude $\varepsilon_{s}$
and the frequency $\omega_s$, is applied on the
system. The dynamics of the
optomechanical system can thus be
described in the rotating frame at the pump
frequency $\omega_p$
\begin{align}
\frac{\mathrm{d}a_1}{\mathrm{d}t} &= i \Delta_1  a_1 -\frac{\kappa_1}{2}a_1 - i g_1 x a_1-\frac{\sqrt{\kappa_{ex1}\kappa_{ex2}}}{2} a_2\nonumber\\ &+\sqrt{\kappa_{ex1}} \varepsilon_p +\sqrt{\kappa_{ex1}} \varepsilon_s e^{-i\delta t}+\sqrt{\kappa_{ex1}}\xi_1 \label{2}\\
\frac{\mathrm{d}a_2}{\mathrm{d}t} &= i \Delta_2  a_2 -\frac{\kappa_2}{2}a_2 - i g_2 x a_2-\frac{\sqrt{\kappa_{ex1}\kappa_{ex2}}}{2} a_1\nonumber\\ &+\sqrt{\kappa_{ex2}} \varepsilon_p +\sqrt{\kappa_{ex2}} \varepsilon_s e^{-i\delta t} +\sqrt{\kappa_{ex2}}\xi_2 \label{3}\\
\frac{\mathrm{d}x}{\mathrm{d}t} &=\frac{P}{m} \label{4}\\
\frac{\mathrm{d}P}{\mathrm{d}t} &= -m \omega^2_m x -\hbar( g_1 a^\dagger_1a_1+ g_2 a^\dagger_2a_2)
-\frac{\Gamma_m}{2} P +\delta F \label{5}
\end{align}
where $\Delta_1 = \omega_p-\omega_1$ and $\Delta_2 = \omega_p-\omega_2$ represent the detuings of optical modes with respect to the driving field. $\kappa_j = \kappa_{j0}+
\kappa_{exj}$ indicates the total
decay rate of optical mode $a_j$ (j=1,2). $\delta = \omega_s-\omega_p$ is the frequency detuning between the probe field and the control field.
And $\xi_1$ ($\xi_2$) is the external noise of the optical mode $a_1$ ($a_2$) introduced by the waveguide. $\delta F$ is the thermal noise
of the mechanical oscillator and $\Gamma_m$ is the decay rate of the mechanical mode. In our case, the pump laser
field is much stronger than the probe field. By using the linearization approach,  thus the Heisenberg operators can be divided into the steady parts and the fluctuation ones, i.e.,
$a_j = \overline{a}_j + \delta a_j$ (j = 1,2) and $x = \overline{x} + \delta x$.
Substituting the division forms into Eq.(\ref{2})-(\ref{5}), the steady solutions of the above dynamical equations can be obtained as
\begin{align}
\overline{a}_1 &= \frac{-\sqrt{\kappa_{ex1}}-\frac{\sqrt{\kappa_{ex1}}\kappa_{ex2}}{2(i\Delta_2-ig_2\overline{x}-\kappa_2/2)}}{i\Delta_1-ig_1\overline{x}-\kappa_1/2-\frac{\kappa_{ex1}\kappa_{ex2}}{4(i\Delta_2-ig_2\overline{x}-\kappa_2/2)}}\varepsilon_p \label{6}\\
\overline{a}_2 &= \frac{\frac{ \sqrt{\kappa_{ex1}\kappa_{ex2}}}{2}\overline{a}_1 -\sqrt{\kappa_{ex2}}\varepsilon_p}{i\Delta_2-ig_2\overline{x}-\kappa_2/2} \label{7}\\
\overline{x} &= -\hbar\frac{g_1|\overline{a}_1|^2+g_2|\overline{a}_2|^2}{m\omega^2_m} \label{8}
\end{align}
Then, we only keep the first-order terms in the small  fluctuation ones $\delta a_1$, $\delta a^\dagger_1$, $\delta a_2$, $\delta a^\dagger_2$ and $\delta x$. Under these conditions, we can obtain the linearized Langevin equations as follows:
\begin{align}
\frac{\mathrm{d}\delta a_1}{\mathrm{d}t} & = (i\overline{\Delta}_1-\frac{\kappa_1}{2}) \delta a_1-ig_1\overline{a}_1 \delta x-\frac{\sqrt{\kappa_{ex1}\kappa_{ex2}}}{2} \delta a_2\nonumber\\
&+\sqrt{\kappa_{ex1}}\varepsilon_s e^{-i\delta t}+\sqrt{\kappa_{ex1}}\xi_1\label{9}\\
\frac{\mathrm{d}\delta a^\dagger_1}{\mathrm{d}t} & = (-i\overline{\Delta}_1-\frac{\kappa_1}{2}) \delta a^\dagger_1+ig_1\overline{a}^*_1 \delta x-\frac{\sqrt{\kappa_{ex1}\kappa_{ex2}}}{2} \delta a^\dagger_2\nonumber\\
&+\sqrt{\kappa_{ex1}}\varepsilon^*_s e^{i\delta t} +\sqrt{\kappa_{ex1}}\xi^\dagger_1\label{10}\\
\frac{\mathrm{d}\delta a_2}{\mathrm{d}t} & = (i\overline{\Delta}_2-\frac{\kappa_2}{2}) \delta a_2-ig_2\overline{a}_2 \delta x-\frac{\sqrt{\kappa_{ex1}\kappa_{ex2}}}{2} \delta a_1\nonumber\\
&+\sqrt{\kappa_{ex2}}\varepsilon_s e^{-i\delta t}+\sqrt{\kappa_{ex2}}\xi_2 \label{11}\\
\frac{\mathrm{d}\delta a^\dagger_2}{\mathrm{d}t} & = (-i\overline{\Delta}_2-\frac{\kappa_2}{2}) \delta a^\dagger_2+ig_2\overline{a}^*_2 \delta x-\frac{\sqrt{\kappa_{ex1}\kappa_{ex2}}}{2} \delta a^\dagger_1\nonumber\\
&+\sqrt{\kappa_{ex2}}\varepsilon^*_s e^{i\delta t} +\sqrt{\kappa_{ex2}}\xi^\dagger_2\label{12}\\
m\frac{\mathrm{d^2}}{\mathrm{d}t^2}\delta x & = -\frac{m\Gamma_m}{2} \frac{\mathrm{d}}{\mathrm{d}t} \delta x-m\omega^2_m \delta x-\hbar g_1\overline{a}_1  (\delta a_1+\delta a^\dagger_1) \nonumber\\
&-\hbar g_2\overline{a}_2 (\delta a_2+\delta a^\dagger_2)+\delta F \label{13}
\end{align}
where $\overline{\Delta}_1=\Delta_1-g_1\overline{x}$ and $\overline{\Delta}_2=\Delta_2-g_2\overline{x}$ denote the effective detuning between the cavity modes and the control laser beam, including
the frequency shift caused by the mechanical motion. 

Combining the above analysis and Fig.\ref{fig1} (c) and (d), we can better understand the multiple light paths for interference. Fig.\ref{fig1} displays a schematic of quantum interference between different light paths. The photons at the output port could come from four different path. The different paths are:
\textcircled{1} the probe photons excite
the cavity mode $a_1$ and pass to the output port; \textcircled{2} the photons generated by the sideband transition  through the optomechanical interaction in mode $a_1$ are coupled out the cavity through the waveguide; \textcircled{3} the probe field passes through the cavity mode $a_2$ directly. In our case, the system only shows three different pathways when the optomechanical coupling rate $g_2$ can be neglected. Otherwise, the fourth pathway is: \textcircled{4} the photons genetated by anti-Stokes process in the optical mode $a_2$ are coupled out the cavity through the waveguide. Thus the photons in the output beam is the sum of four different paths. For convenience, we introduce $x=x_{ZPF}(b+b^\dagger)$ and $P=-i m \omega_m x_{ZPF} (b-b^\dagger)$ to get the transmission of the system, where $x_{ZPF}$ is the zero point fluctuation. We also neglect the quantum noise $\xi_1$, $\xi_2$ and $\delta F$. Without loss of generality, in the following discussions the frequency of the auxiliary cavity mode is tuned to satisfy $\overline{\Delta}_1 =\overline{\Delta}_2 =\Delta$. According to the input-output theory\cite{gardiner2004quantum}, we can
get the output of the probe field:
$a_{out} = \varepsilon_s-\sqrt{\kappa_{ex1}}
a_1-\sqrt{\kappa_{ex2}} a_2$.
With neglecting
the high order sideband effect, the normalized transmission coefficient in the red sideband driving regime can be simplified to 
\begin{align}
t_r & = \frac{(L_1-\kappa_{ex1})(L_2-\kappa_{ex2})-(A-\sqrt{\kappa_{ex1}\kappa_{ex2}})^2}{L_2(L_1-\frac{A^2}{L_2})}
\end{align}
where $G_1$ and $G_2$ are the effective optomechanical coupling strength, $L_1=i(-\Delta-\delta)+\frac{\kappa_1}{2}+\frac{G^2_1}{i(-\Delta-\delta)+\frac{\Gamma_m}{2}}$, $L_2=i(-\Delta-\delta)+\frac{\kappa_2}{2}+\frac{G^2_2}{i(-\Delta-\delta)+\frac{\Gamma_m}{2}}$ and $A=\frac{G_1 G_2 }{i(-\Delta-\delta)+\frac{\Gamma_m}{2}}+\frac{\sqrt{\kappa_{ex1}\kappa_{ex2}}}{2}$. 
When the system is driven by the blue-detuned pump field, the normalized transmission coefficient is
\begin{align}
t_b & =\frac{(R_1+\kappa_{ex1})(R_2+\kappa_{ex2})-(B-\sqrt{\kappa_{ex1}\kappa_{ex2}})^2}{R_2(R_1-\frac{B^2}{R_2})}
\end{align}
where $R_1=i(\Delta+\delta)-\frac{\kappa_1}{2}-\frac{G^2_1}{i(\Delta+\delta)-\frac{\Gamma_m}{2}}$, $R_2=i(\Delta+\delta)-\frac{\kappa_2}{2}-\frac{G^2_2}{i(\Delta+\delta)-\frac{\Gamma_m}{2}}$ and $B=\frac{G_1 G_2 }{i(\Delta+\delta)-\frac{\Gamma_m}{2}}+\frac{\sqrt{\kappa_{ex1}\kappa_{ex2}}}{2}$. And the corresponding power transmission coefficient is
given by $T = |t|^2$
\begin{figure}[htbp]
\centering
\includegraphics[width=0.9\linewidth]{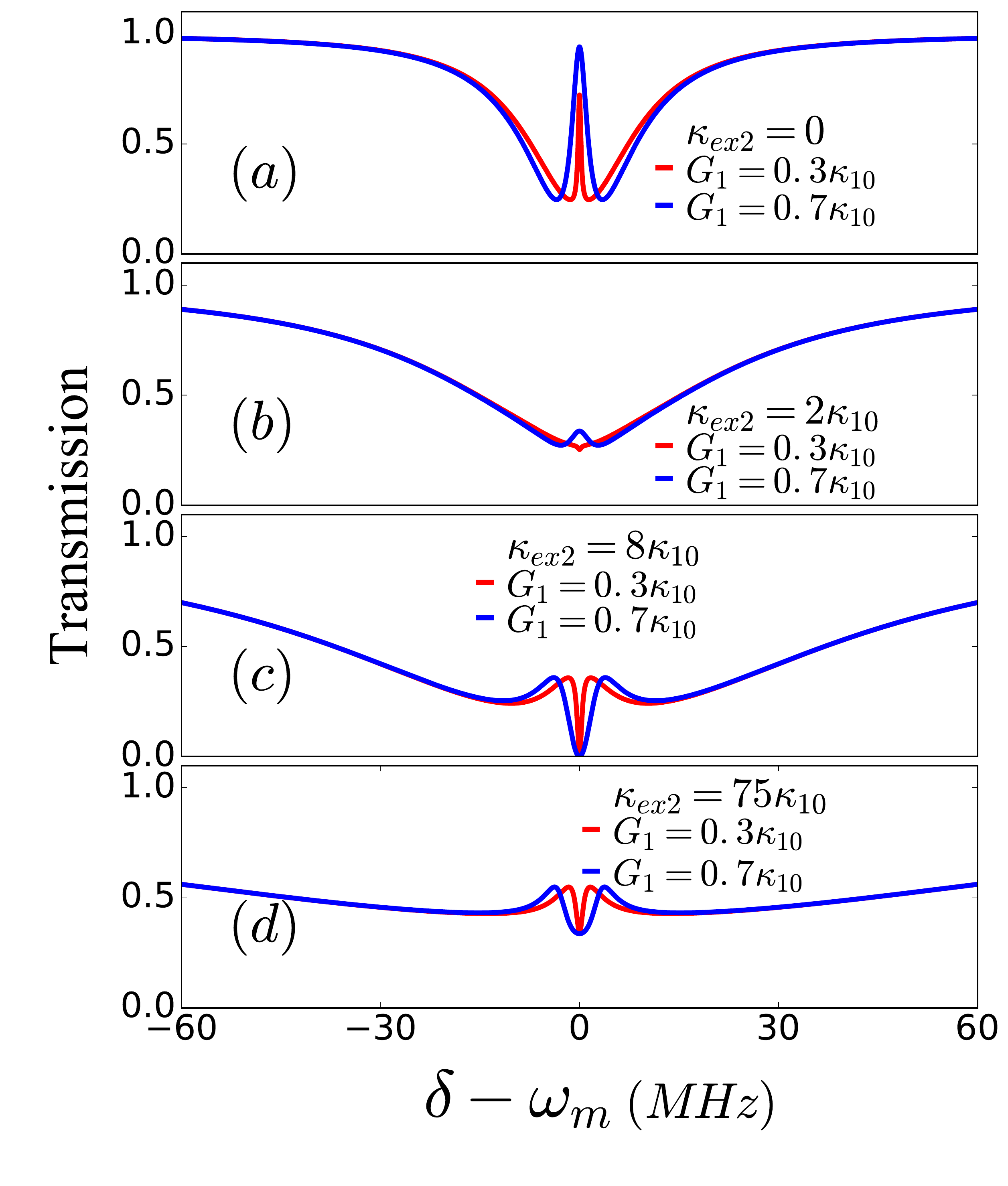}
\caption{The transmission rate are plotted with the different external coupling strength $\kappa_{ex2}$ of the auxiliary cavity mode $a_2$ in the red sideband regime. The parameters used here are: $\kappa_{10}=5$ MHz, $\kappa_{ex1}=15$ MHz, $\kappa_{20}=40$ MHz,  $\Gamma_m=50$ KHz, $\omega_m=100$ MHz and $\Delta = -\omega_m$.}
\label{fig2}
\end{figure}
\section{Transmission rate with three pathways interference}
\label{section:3}
Before we discuss the three-pathway interference
effect, the two-pathway induced
EIT-like effect is considered. The all-optical
analogues of the EIT effect are widely studied in
optical resonators systems. And this EIT-like effect contains two coupling mechanism: directly
coupled-resonator-induced transparency(DCRIT) \cite{totsuka2007slow,yanik2004stopping} and indirectly
coupled-resonator-induced transparency(ICRIT) \cite{xiao2009electromagnetically,dong2009modified}. By combining the DCRIT and OMIT
effect, the three-pathway induced EIT and EIA effects
has been studied in different system \cite{lei2015three,bai2016tunable,si2017optomechanically,zhang2017optomechanically,liu2017controllable}. For the coupled cavity system, the two cavities should be tuned precisely to couple, requiring the complicated control and fabrication for the expereimtns. Thus here we study
the three-pathway interference effect by combining
the ICRIT effect and optomechanical interaction in a single
cavity. In this section, we consider the situation that the auxiliary cavity mode $a_2$ decouples to the mechanical mode, i.e., $g_2=0$. We study the power transmission coefficient with the effect of the auxiliary cavity mode both in the red and blue sideband driving regime. We find that the three mode interaction could be tuned by mode $a_2$, which can be used to control the transmission rate.
\subsection{Three-pathway interference with red-detuned driving}
\label{subsection:1}
We have discussed the physical origins of the three pathway interference phenomena in Fig.\ref{fig1} (c). The photons at the output port with frequency equaling to the probe signal come from the three pathways. When one light path is changed, the relative amplitude and phase of the multiple light paths interference is adjusted. This influence of the three-pathway interference effect will change the output spectrum. To investigate the influence of path \textcircled{3} on the three-pathway interference effect, we plot the output spectrum in Fig.\ref{fig2} with different external coupling rate $\kappa_{ex2}$. In Fig.\ref{fig2} (a), the transmission shows the OMIT window without the auxiliary mode. This is the result of the two-pathway destructive interference effect. When the optical mode $a_2$ couples with tapered fiber, the transparent peak maintains but becomes really  minute plotted by blue line (@ $G_1 / \kappa_{10} =0.7$) in (b). While the absorption dip starts appearing represent by the red solid line with the weaker optomechanical coupling rate $G_1=1.5$ MHz. 
\begin{figure}[htbp]
\centering
\includegraphics[width=\linewidth]{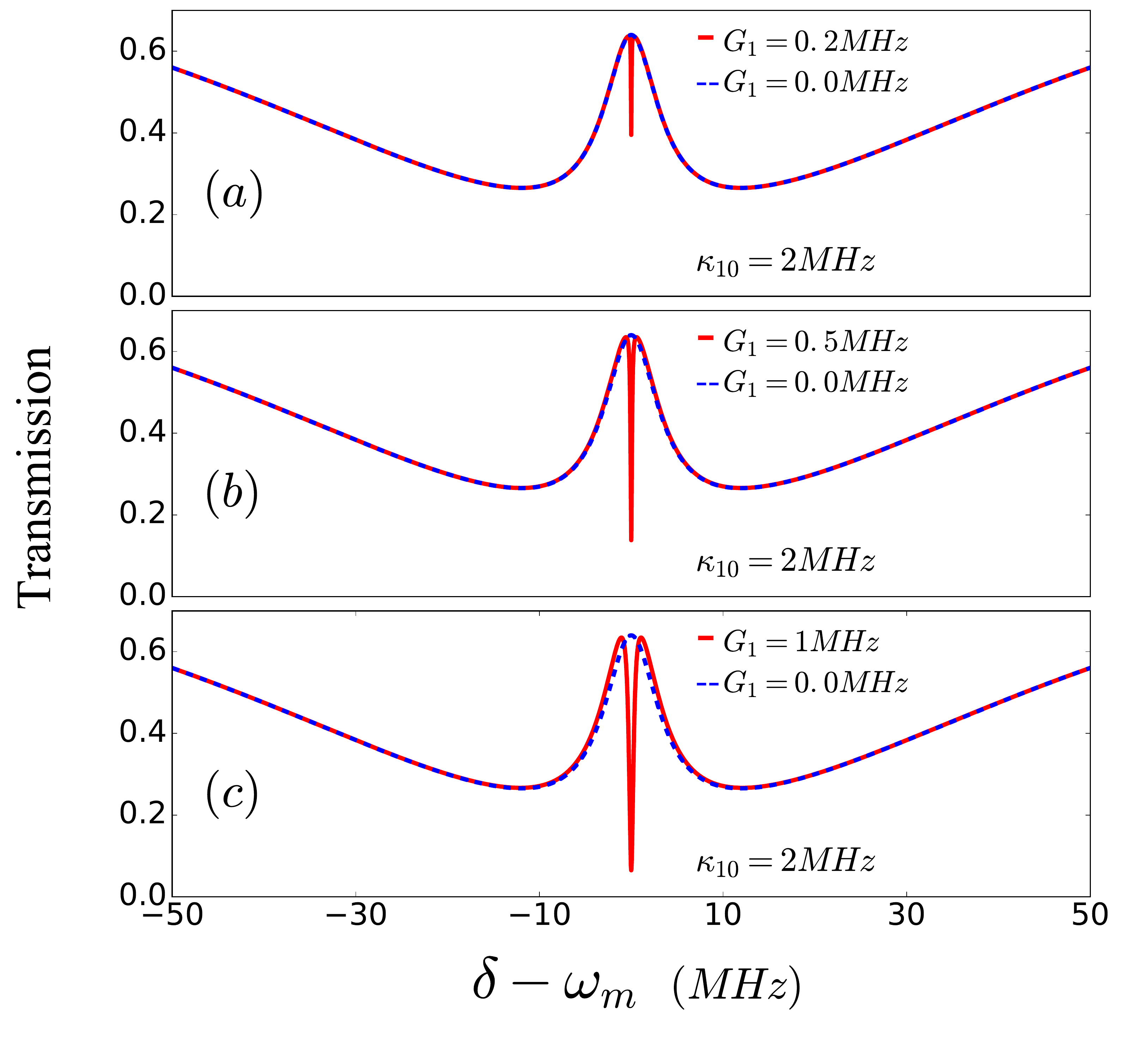}
\caption{The transmission rate are plotted with the different optomechanical coupling rate between the optical modes and the phonons. The parameters used here are: $\kappa_{ex1}=15$ MHz, $\kappa_{20}=40$ MHz,  $\kappa_{ex2}=60$ MHz,  $\Gamma_m=50$ KHz, $\omega_m=100$ MHz  and $\Delta = -\omega_m$.}
\label{fig3}
\end{figure}
Further increasing the external coupling rate $\kappa_{ex2}$ of the auxiliary mode to 40 MHz in Fig.\ref{fig2} (c), the output spectra show the evident absorption dips within a EIT-like window for both lines. The power transmission drops to zero due to the constructive interference. By tuning the external coupling rate $\kappa_{ex2}/\kappa_{10}$ to $75$, the tapered fiber mediated interference
\begin{figure}[htbp]
\centering
\includegraphics[width=\linewidth]{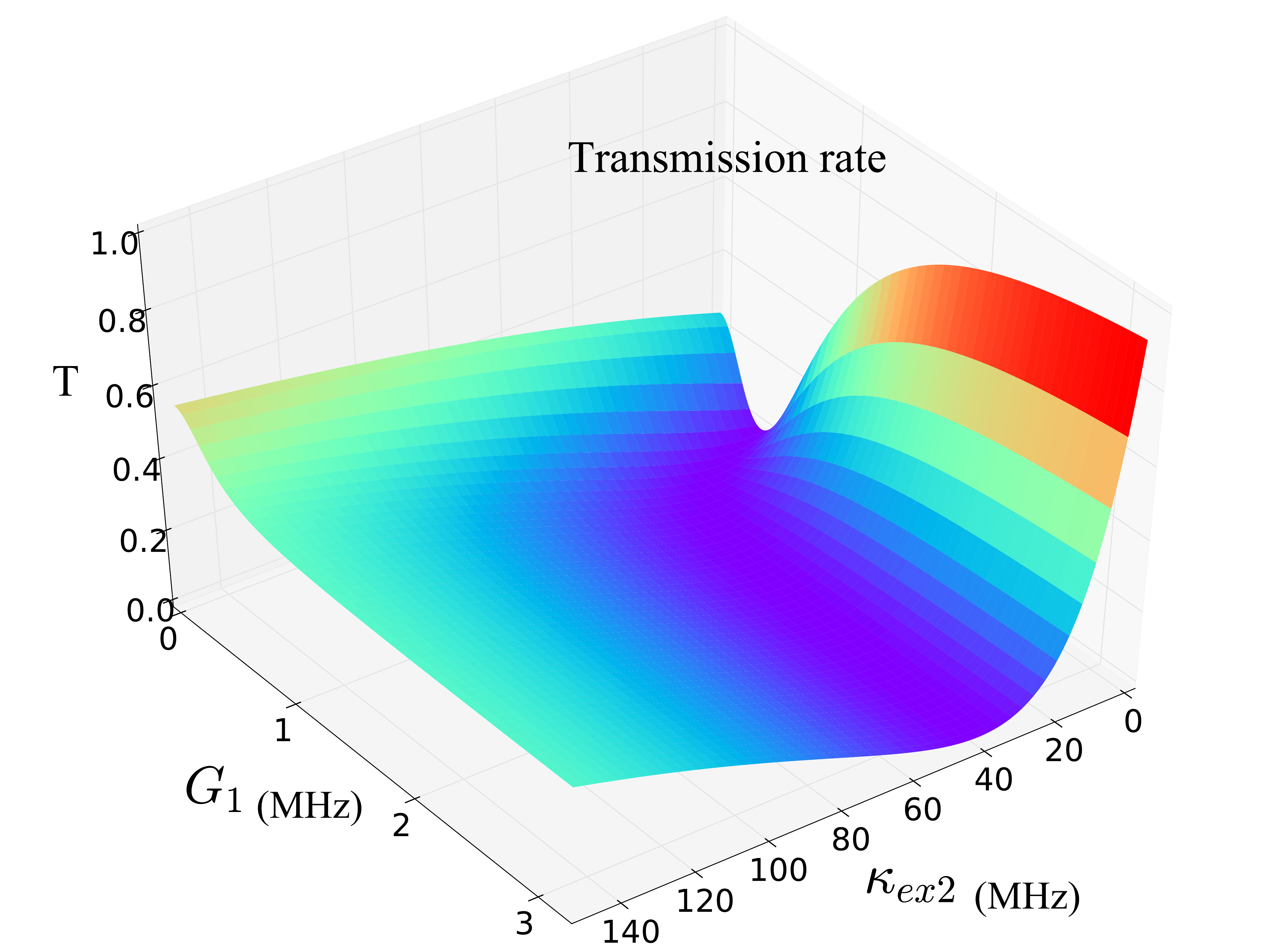}
\caption{The transmission is plotted as the function of the optomechanical coupling rate $G_1$ and the external coupling strength $\kappa_{ex2}$ of the auxiliary cavity mode. The parameters used here are: $\kappa_{10}=5$ MHz, $\kappa_{ex1}=15$ MHz, $\kappa_{20}=40$ MHz,   $\Gamma_m=50$ KHz, $\omega_m=100$ MHz and $\Delta = -\omega_m$.}
\label{fig4}
\end{figure}
\begin{figure*}[htbp]
\centering
\includegraphics[width=0.9\linewidth]{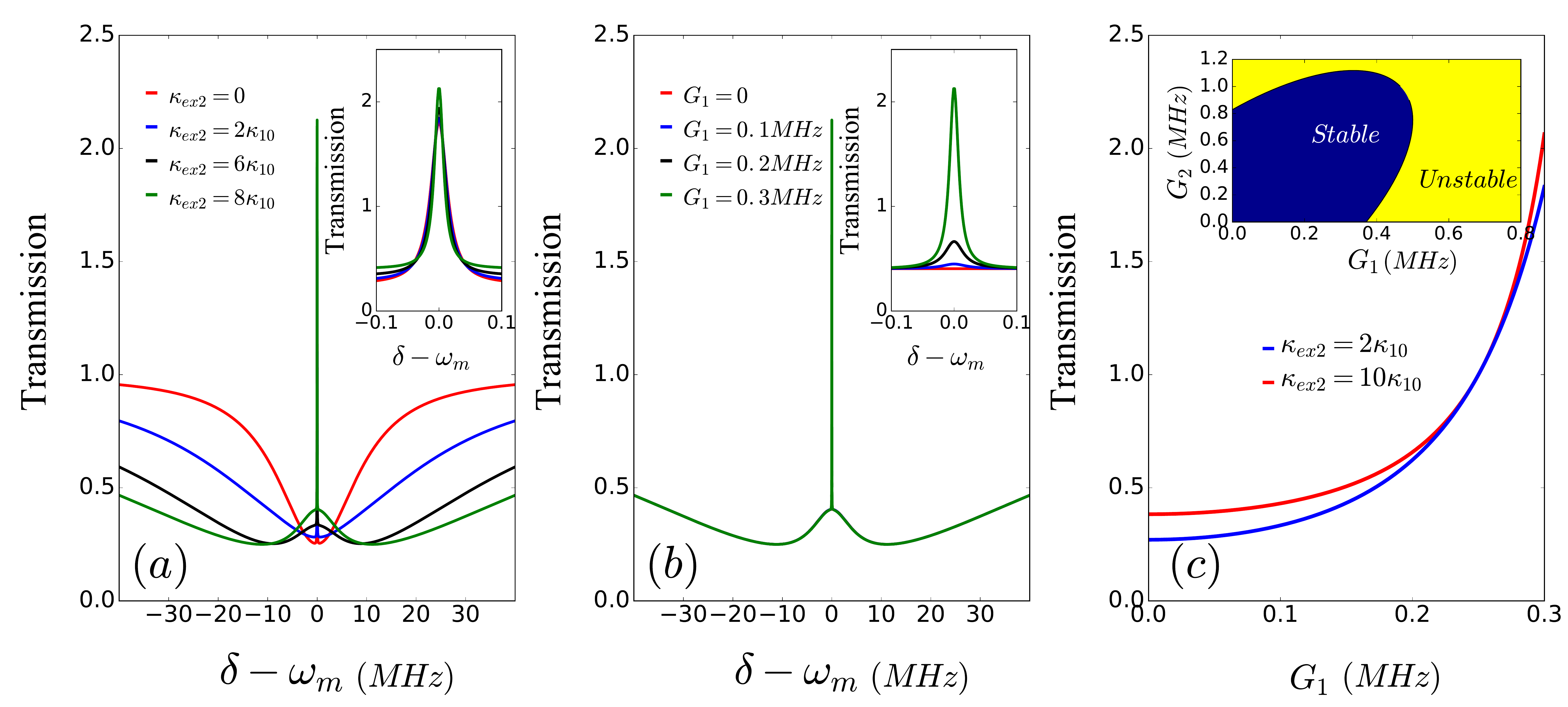}
\caption{The transmission rate are plotted in the blue sideband driving regime when the auxiliary cavity mode decouples to the mechanical mode. (a) The external coupling rate $\kappa_{ex2}$ varies from 0 to $8\kappa_{10}$ when $G_1$ is fixed to 0.3 MHz. (b) Here $\kappa_{ex2}$ is fixed to 60 MHz while the $G_1$ varies from 0 to 0.3 MHz. (c) The transmission of the central peak is presented as the function of the optomechanical coupling strength $G_1$. The parameters used here are: $\kappa_{10}=5$ MHz, $\kappa_{ex1}=20$ MHz, $\kappa_{20}=40$ MHz, $\Delta=\omega_m$, $\Gamma_m=50 $ KHz and $G_2 = 0$. The inset of figure (c) shows the border between the stable and unstable area when $\kappa_{ex2}= 60$ MHz.  }
\label{fig5}
\end{figure*}
effect is strong compared with the optomechanical coupling strength. Both the transmission spectrum and the absorption window become wider. While the transmission ($\delta=\omega_m$) increases to 0.33 even with the absorption dip.

Form Fig.\ref{fig2}, we can obtain that the interference between the two optical modes could lead the transmission
rate form OMIT peak to OMIA dip. The reason for the absorption windows is the constructive interference effect induced by the auxiliary cavity mode. Without the auxiliary cavity mode $a_2$, the output field only connects with mode $a_1$ and shows a OMIT peak in  Fig.\ref{fig2} (a). When the auxiliary cavity mode is introduced, the photons with frequency equaling to the probe signal come from the optical modes $a_1$ and $a_2$ both. While the anti-stokes process leads to destructive interference for the intracavity field $a_1$, suppressing the photons population of $a_1$. The constructive interference leads an absorption dip and the transmission rate of central dip drops to zero. 

Then we study the influence of the optomechanical interaction on the three-pathway interference. The transmission rate is plotted in Fig.\ref{fig3} with the different optomechanical coupling strength $G_1$. In Fig.\ref{fig3} (a), (b) and (c), the external coupling strength $\kappa_{ex2}$ is fixed to 60 MHz and $G_2=0$. The transmission displays the EIT-like window with $G_1=0$ plotted by the blue dashed lines. When the optomechanical coupling strength $G_1$ is increased to 0.4 MHz , there is an small absorption window within the EIT window in Fig.\ref{fig3} (a). In Fig.\ref{fig3}(c), the transmission of the central dip drops to 0.1 with stronger optomechanical coupling rate $G_1$. It is obvious that with increasing the optomechanical
coupling rate, the transmission shows more prominent absorption effect.  

From the Fig.\ref{fig2} and Fig.\ref{fig3}, we can obtain that the three-pathway interference effect could vary from the transparent window to an absorption
dip in our system. The output spectrum 
is sensitive to the auxiliary cavity mode in Fig.\ref{fig2}, which can be used to
control the three mode interaction. While in Fig.\ref{fig3}, the depth of the absorption dip can be manipilated by the path \textcircled{2}.  This system offers two effective methods to tune the optical response. The transmission is plotted in Fig.\ref{fig4} as the function of the optomechanical coupling rate $G_1$ and the external coupling strength $\kappa_{ex2}$ of the auxiliary cavity mode. It can be obtained that with the weak optomechanical coupling rate, the effect of two optical modes interference becomes more evident and leads to transparency-like window. While the transmission shows OMIT window when $\kappa_{ex2}$ is weak.
Thus the transmission can be manipulated through tuning the relative amplitudes and phases of the three-pathway interference.
\begin{figure*}[htbp]
\centering
\includegraphics[width=\linewidth]{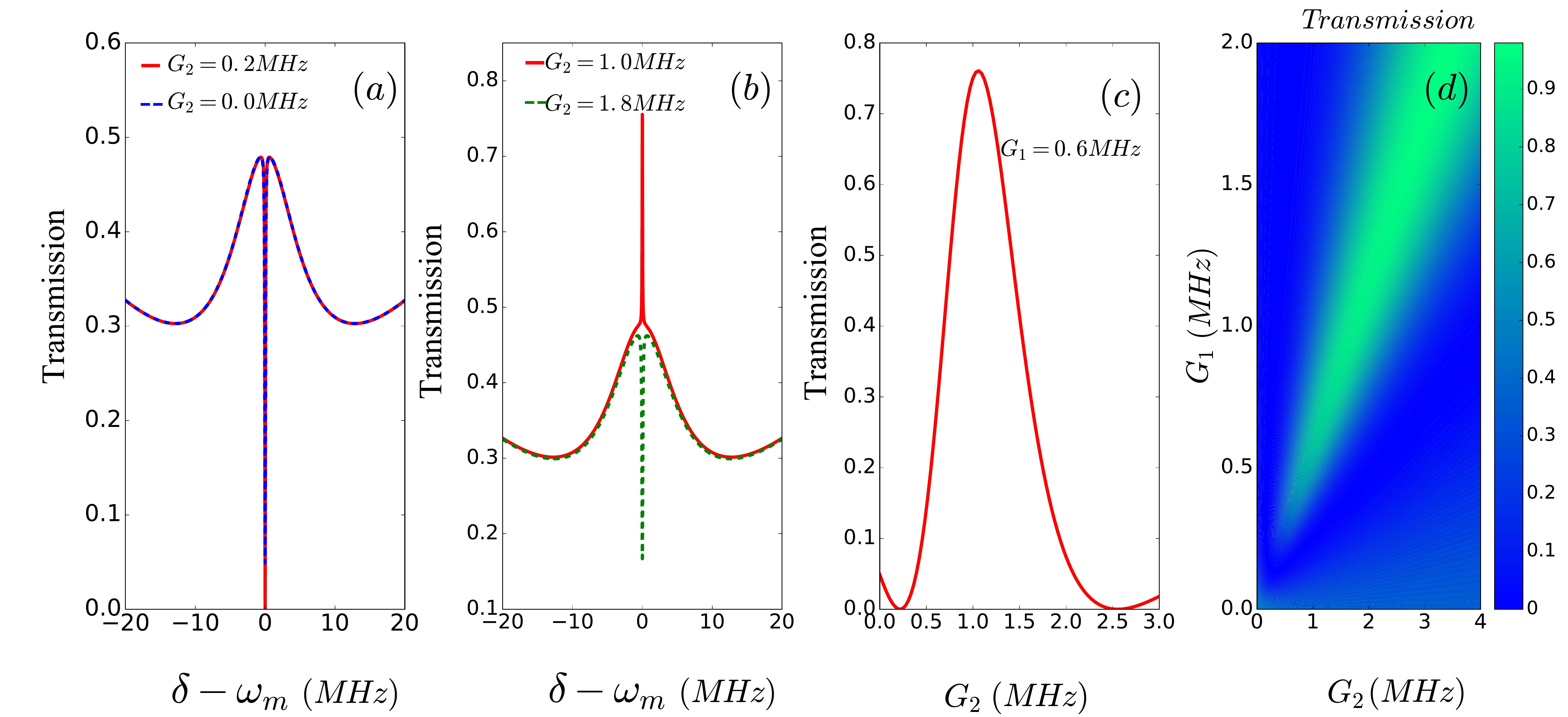}
\caption{The output spectra are shown in (a) and (b) as a function of $\delta$ with different $G_2$. The effective optomechanical coupling strength $G_1$ is fixed to 0.6 MHz. The central peak is plotted in (c) as a function of $G_2$. The power transmission coefficient
T as a function of $G_2$ and $G_1$ in (d). Here $\kappa_{10}=5$ MHz, $\kappa_{ex1}=20$ MHz, $\kappa_{20}=40$ MHz, $\kappa_{ex2}=60$ MHz,  $\Gamma_m=5$ KHz and $\Delta = -\omega_m$.}
\label{fig6}
\end{figure*}
\subsection{Three-pathway induced amplification with blue-detuned driving}
\label{subsection:2}
In our system, the photons generated form the optomechanical interaction will directly affect the intracavity probe photons population, which could result in OMIA dip in the red sideband pumping regime. Now we consider the situation that the system is driven by a blue-detuned driving field and the auxiliary optical mode decouples to the mechanical mode. In this scheme we limit our parameters to the stable driving regime. And the stability condition can be obtained by
analyzing the Lyapunov exponents \cite{benettin1980lyapunov} of the Jacobian
matrix. We have plotted the border between the stable and unstable condition in the inset of Fig.\ref{fig5}(c). In
Fig.\ref{fig5}, we have shown the the normalized output spectra of optomechanical system with a blue-detuned pump field. In Fig.\ref{fig5} (a), we plot the transmission rate as a function of $\delta$ for different values of external coupling rate $\kappa_{ex2}$. The red solid line indicates the transmission with optomechanical coupling rate $G_1=0.3$ MHz while $\kappa_{ex2}=0$.
Without the auxiliary optical mode, it is clear that the transmission line shows
an amplified peak. With the
enhancement of the two optical mode
interference through increasing
$\kappa_{ex2}$, the transmission becomes broader and shows the EIT-like window. As the inset of Fig.\ref{fig5} (a) shows, the transmission
at the central peak could enhanced with stronger external coupling strength of the auxiliary optical mode. 

We also plot the transmission amplification effect for different values of the coupling strength $G_1$ in Fig.\ref{fig5}
(b). Here $\kappa_{ex2}$ is fixed to 60 MHz,
and $G_1$ is increased to 0.3 MHz. The red solid
line shows a EIT-like window without optomechanical interaction. It is clear that the transmission rate of central peak increases to 1.75 with $G_1=0.3$ MHz compared with red line. In Fig.\ref{fig5}(b), the transmission amplification effect of the probe field 
becomes stronger with increasing the
optomechanical coupling rate $G_1$. Fig.\ref{fig5} (c) plots the transmission rate of the central peak as a function of the coupling rate $G_1$ when the probe frequency detuning
$\delta =\omega_m$. The blue (red) solid line plots the amplification effect
when $\kappa_{ex2}$ equals to 10 (60) MHz. When $G_1$ is small, the fiber mediated two optical interference is strong, the transmission rate depends on $\kappa_{ex2}$. For strong optomechanical coupling rate, the amplification depends on $G_1$, while the transmission can be affected by the auxiliary cavity mode.   

By increasing the optomechancial coupling rate $G_1$, the stokes process is enhanced and emits photons and phonons. The photons generated by the optomechanical interaction are degenerate with the probe field, which accelerates the circulating power of $a_1$. The two optical interference is weak compared with the optomechanical coupling. Thus the optomechanical amplification plays a dominant role in the three pathways. The closer to the stability border, the more prominent the optomechanical gain effect is. However, even the central peak is not sensitive to $\kappa_{ex2}$, the amplification effect could be enhanced by the two optical interference.
\section{Transmission rate with four-pathway interference effect}
\label{section:4}
In this section, we will consider the situation that the auxiliary cavity mode couples to the mechanical mode. Here we assume that the effective optomechanical coupling strength $G_1$ and $G_2$ can be tuned separately by using the extra tapered fiber to introduce the extra pump field.

To investigate the effect of optomechanical coupling rate $G_1$, $G_2$ on the transmission spectrum, we plot the transmission with the different optomechanical couping rates $G_1$ and $G_2$ in Fig.\ref{fig6} with the red-detuned driving. Under the three pathways situation in red sideband regime, the transmission rates show an absorption window, which have been presented in Fig.\ref{fig2}. Through increasing the optomechanical coupling rate $G_1$, the absorption dip becomes deeper and wider. Different with three pathways induced absorption, four pathways interference effect can switch the transmission spectrum back and forth between OMIT and OMIA depending on the coupling strength $G_1$ and $G_2$. In Fig.\ref{fig6}(a), when $G_2=0$, the transmission of the central dip is 0.13 which has been presented by the blue dashed line. While the absorption dip becomes more evident in the middle of transmission window
when $G_1=0.6$ MHz and $G_2=0.2$ MHz. The transmission drops to 0 due to constructive interference effect. By keeping $G_1 =0.6$ MHz unchanged, the transmission rate is plotted in Fig.\ref{fig6}(b) when the coupling rate $G_2$ is enlarged to 1 MHz. In contrast to the absorption dip for the coupling rate $G_2=0.2$ MHz, here the OMIT peak appears for the resonant case, making the transmission go up to 0.75. By increasing the coupling rate $G_2$ to 1.8 MHz, the constructive interference between different pathways results in the OMIA again. To show the manipulation between OMIA and OMIT through tuning the ratio between $G_1$ and $G2$, the height of the central peak is plotted in Fig.\ref{fig6}(c). It is clear that when the optomechanical coupling strength $G_2$ is weak, the four-pathway interference will enhance the constructive
interference and lead to the OMIA effect. Further increasing $G_2$, the four pathways will result in the destructive interference and OMIT window. While $G_2$ is larger than 1.5 MHz, the constructive interference occurs, resulting in the absorption window again. In Fig.\ref{fig6}(d), we plot the transmission rate as a function of different coupling rate $G_1$ and $G_2$. We can obtain that the conversion between the constructive and destructive interference can be achieved by tuning the coupling strength of $G_1$ and $G_2$.  Compared with $T=0.51$ when the coupling rate $G_1 = G_2 = 0$, transmission can be switched to 0 or 1 due to the OMIA or OMIT effect. 
\section{Conclusion}
\label{section:5}
In summary, we have explored an optomechanical system with multiple light paths interference effect in a single optical resonator. We give the explicit physical explanations and detailed calculations in this paper. By combining the EIT-like and OMIT effect, the system shows destructive or constructive interference effect under different conditions. The auxiliary cavity mode offers the additional light pathway to control the output spectrum. Through enhancing the optomechanical coupling strength or the external coupling rate of the auxiliary cavity mode, we can tune the optical response of the probe field effectively. Experimentally, the system has no limitation on the quality factor of the auxiliary cavity mode, making it easy to implement. Moreover, our model paves an easy way for realization of multiple
interference and manipulation of OMIT and OMIA. 
 \section*{Acknowledgments} 
The work was supported by the National Natural Science Foundation of
China (NSFC) (20171311628); Ministry of Science and
Technology of the People’s Republic of China (MOST)
(2017YFA0303700); Beijing Advanced Innovation Center
for Future Chip (ICFC).

\bibliographystyle{unsrt}
\bibliography{main}
\end{document}